\documentclass[%
 reprint,
 superscriptaddress,
 amsmath,amssymb,
 aps,
 prb,
]{revtex4-2}

\setcitestyle{number}
\usepackage{graphicx}% Include figure files
\usepackage{amsmath,amsfonts,amssymb}
\usepackage[dvipsnames]{xcolor}
\usepackage{dcolumn}% Align table columns on decimal point
\usepackage{bm}% bold math
\usepackage{hyperref}% add hypertext capabilities
\hypersetup{
    colorlinks=true,
    linkcolor=blue,
    filecolor=magenta,      
    urlcolor=blue,
    citecolor=blue
    }
\usepackage{siunitx}

\begin{document}

\title{Spectroscopic Signatures of Structural Disorder and Electron-Phonon Interactions in Trigonal Selenium Thin Films for Solar Energy Harvesting}

\author{Rasmus S. Nielsen}
\email[]{Electronic mail: rasmus.nielsen@empa.ch}
\affiliation{Nanomaterials Spectroscopy and Imaging, Transport at Nanoscale Interfaces Laboratory, Swiss Federal Laboratories for Material Science and Technology (EMPA), Ueberlandstrasse 129, 8600 Duebendorf, Switzerland}

\author{Axel G. Medaille}
\affiliation{Universitat Politècnica de Catalunya (UPC), Photovoltaic Lab - Micro and Nano Technologies Group (MNT), Electronic Engineering Department, EEBE, Av Eduard Maristany 10-14, Barcelona 08019, Spain}
\affiliation{Universitat Politècnica de Catalunya (UPC), Barcelona Centre for Multiscale Science \& Engineering, Av Eduard Maristany 10-14, Barcelona 08019, Spain}

\author{Arnau Torrens}
\affiliation{Universitat Politècnica de Catalunya (UPC), Photovoltaic Lab - Micro and Nano Technologies Group (MNT), Electronic Engineering Department, EEBE, Av Eduard Maristany 10-14, Barcelona 08019, Spain}
\affiliation{Universitat Politècnica de Catalunya (UPC), Barcelona Centre for Multiscale Science \& Engineering, Av Eduard Maristany 10-14, Barcelona 08019, Spain}

\author{Oriol Segura-Blanch}
\affiliation{Universitat Politècnica de Catalunya (UPC), Photovoltaic Lab - Micro and Nano Technologies Group (MNT), Electronic Engineering Department, EEBE, Av Eduard Maristany 10-14, Barcelona 08019, Spain}
\affiliation{Universitat Politècnica de Catalunya (UPC), Barcelona Centre for Multiscale Science \& Engineering, Av Eduard Maristany 10-14, Barcelona 08019, Spain}

\author{Seán R. Kavanagh}
\affiliation{Harvard University Center for the Environment, Cambridge, Massachusetts 02138, United States}

\author{David O. Scanlon}
\affiliation{School of Chemistry, University of Birmingham, Birmingham B15 2TT, UK}

\author{Aron Walsh}
\affiliation{Thomas Young Centre and Department of Materials, Imperial College London, London SW7 2AZ, UK}
\affiliation{Department of Physics, Ewha Womans University, Seoul 03760, Korea}

\author{Edgardo Saucedo}
\affiliation{Universitat Politècnica de Catalunya (UPC), Photovoltaic Lab - Micro and Nano Technologies Group (MNT), Electronic Engineering Department, EEBE, Av Eduard Maristany 10-14, Barcelona 08019, Spain}
\affiliation{Universitat Politècnica de Catalunya (UPC), Barcelona Centre for Multiscale Science \& Engineering, Av Eduard Maristany 10-14, Barcelona 08019, Spain}

\author{Marcel Placidi}
\affiliation{Universitat Politècnica de Catalunya (UPC), Photovoltaic Lab - Micro and Nano Technologies Group (MNT), Electronic Engineering Department, EEBE, Av Eduard Maristany 10-14, Barcelona 08019, Spain}
\affiliation{Universitat Politècnica de Catalunya (UPC), Barcelona Centre for Multiscale Science \& Engineering, Av Eduard Maristany 10-14, Barcelona 08019, Spain}

\author{Mirjana Dimitrievska}
\email[]{Electronic mail: mirjana.dimitrievska@empa.ch}
\affiliation{Nanomaterials Spectroscopy and Imaging, Transport at Nanoscale Interfaces Laboratory, Swiss Federal Laboratories for Material Science and Technology (EMPA), Ueberlandstrasse 129, 8600 Duebendorf, Switzerland}

\begin{abstract} 

Selenium is experiencing renewed interest as a elemental semiconductor for a range of optoelectronic and energy applications due to its irresistibly simple composition and favorable wide bandgap. However, its high volatility and low radiative efficiency make it challenging to assess structural and optoelectronic quality, calling for advanced, non-destructive characterization methods. In this work, we employ a closed-space encapsulation strategy to prevent degradation during measurement and enable sensitive probing of vibrational and optoelectronic properties. Using temperature-dependent Raman and photoluminescence spectroscopy, we investigate grown-in stress, vibrational dynamics, and electron-phonon interactions in selenium thin films synthesized under nominally identical conditions across different laboratories. Our results reveal that short-range structural disorder is not intrinsic to the material, but highly sensitive to subtle processing variations, which strongly influence electron-phonon coupling and non-radiative recombination. We find that such structural disorder and grown-in stress likely promote the formation of extended defects, which act as dominant non-radiative recombination centers limiting carrier lifetime and open-circuit voltage in photovoltaic devices. These findings demonstrate that the optoelectronic quality of selenium thin films can be significantly improved through precise control of synthesis and post-deposition treatments, outlining a clear pathway toward optimizing selenium-based thin film technologies through targeted control of crystallization dynamics and microstructural disorder.

\end{abstract}

\maketitle

\section{Introduction}

\begin{figure*}[t!]
    \centering
    \includegraphics[width=\textwidth,trim={0 0 0 0},clip]{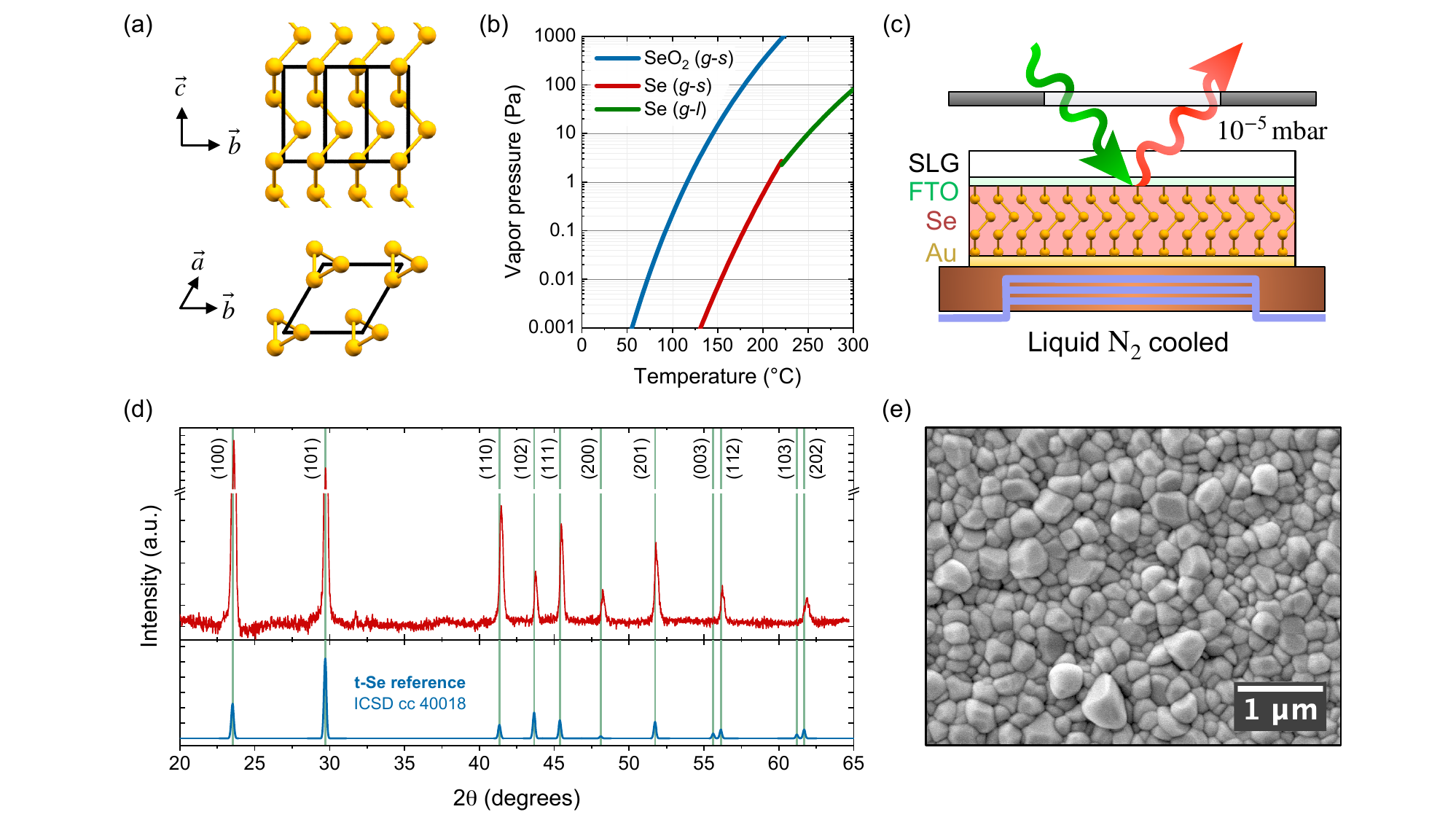}%TRIM=0 0 0 0
    \caption{(a) Crystal structure of trigonal selenium viewed along different lattice directions. (b) Vapor pressure of selenium and selenium dioxide as a function of temperature; \textit{g}, \textit{l}, and \textit{s} denote gas, liquid, and solid phases, respectively. Data from Xu \textit{et al}. \cite{xu2020a} (c) Device schematic of the closed-space encapsulated selenium thin film in an optical cryostat cooled using liquid nitrogen at $\sim\text{10}^\text{-5}$ mbar. (d) XRD pattern of the polycrystalline selenium thin film, shown above the reference powder pattern for trigonal selenium (ICSD collection code 40018); all observed reflections match the reference. (e) Top-view SEM image of the polycrystalline selenium thin film.}
    \label{fig:Figure1}
\end{figure*}

Selenium is a wide-bandgap elemental semiconductor that has attracted renewed interest for a range of optoelectronic applications, including photodetectors \cite{li2024a, chen2024a, adachi2023a}, radiation sensors \cite{kasap2000a, belev2004a}, and, most notably, indoor \cite{yan2022a, wang2024a, wei2023a, bishop2017a, lu2025a, placidi2025a} and tandem photovoltaics \cite{youngman2021a, nielsen2024b, dou2025a, salem2025a, singh2025a}. Its quasi-one-dimensional trigonal crystal structure, shown in Figure \ref{fig:Figure1}(a), gives rise to pronounced anisotropy in material properties, enabling more advanced technologies such as polarization-sensitive light sources and detectors \cite{wang2001a, nanot2013a, yuan2015a, liu2024a}. In addition to its low-dimensional structure, trigonal selenium offers a direct bandgap in the 1.8-2.2 eV range \cite{zhu2019a, bao2025a, hadar2019b}, which is tunable through alloying with tellurium \cite{hadar2019a, zheng2022a, deshmukh2022a}, a high optical absorption coefficient \cite{nielsen2022a, yan2022a}, tolerance to intrinsic point defects \cite{kavanagh2025a, li2024b, moustafa2024a}, low-temperature processing \cite{lu2022a, fu2022a, lu2024a, an2025a}, long-term air stability \cite{zhu2016a, liu2020a}, and the irresistible simplicity of a single-element system. Yet despite these advantages, the optoelectronic quality of selenium thin films remains a major bottleneck, limiting progress toward high-efficiency photovoltaic devices \cite{nielsen2022a, kavanagh2025a}. Photoluminescence (PL) and Raman spectroscopy is widely recognized as a powerful non-contact characterization tools for evaluating optoelectronic and structural quality of materials \cite{kirchartz2020a, nielsen2025c, dimitrievska2016a}. Therefore, a systematic investigation of the radiative emission properties and microstructural quality using these techniques is a natural next step toward identifying both intrinsic and synthesis-induced performance limitations in state-of-the-art selenium thin films.

PL studies of trigonal selenium date back to the 1960s and 1970s, but focus exclusively on bulk crystals grown via vapor phase or melt growth and measured at cryogenic temperatures. Queisser et al. reported polarization-dependent PL from selenium crystals submerged in liquid hydrogen, revealing sharp spectral lines attributed to bound excitons \cite{queisser1967a}. Zetsche et al. observed numerous phonon satellites in PL spectra collected between 2–50 K \cite{zetsche1969a}, and Kuhler et al. performed electro-PL measurements under applied electric fields to probe impurity-related transitions \cite{kuehler1973a}. Later, Moreth identified sharp emission lines from indirect free-exciton decay at 1.4 K, enabling an unambiguous determination of the quasi-indirect nature of the bandgap \cite{moreth1979a}.

In contrast to free-standing single crystals, selenium thin films are typically grown directly on substrates and are thus more susceptible to substrate-induced stress, anisotropic grain growth, and structural disorder -- all of which can significantly alter PL behavior. In 2022, Nielsen et al. reported two distinct PL emission bands at cryogenic temperatures from a selenium thin film integrated into a record-performing photovoltaic device \cite{nielsen2022a}. Liu et al. observed similar spectral features, although the acquisition conditions were not specified \cite{liu2023a}. Other studies have reported redshifted PL peak energies well below the optical bandgap in films infiltrated into mesoporous oxide scaffolds \cite{deshmukh2022a, wu2019a}. Li et al. presented PL spectra with peak energies more reasonable for the optical bandgap, but without a detailed analysis of the nature of the underlying transition \cite{li2024a}. This inconsistency across the literature underscores the need for a more systematic investigation. However, the high vapor pressure of selenium makes thin films prone to sublimation and degradation during PL measurements, even under low excitation conditions. This challenge may have limited the widespread use of PL as a diagnostic tool, despite its value for evaluating optoelectronic quality.

In this work, we use temperature-dependent PL and Raman spectroscopy to study the optoelectronic and vibrational properties of selenium thin films synthesized in different laboratories using state-of-the-art fabrication methodologies. To address the challenge posed by the high vapor pressure and volatility of selenium, we adopt a recently developed closed-space encapsulation strategy introduced by Nielsen et al. \cite{nielsen2024a}, which effectively suppresses sublimation during the measurements. The vibrational properties are studied through a detailed analysis of the thermal evolution of Raman peaks, supported by first-principles phonon dispersion calculations that enable identification of all Raman-active modes and their associated atomic displacements. Temperature-dependent PL measurements, complemented by power-dependent excitation studies, are used to examine the nature of two emission bands and quantify the electron-phonon interactions that drive the thermal shifts and broadening of the observed radiative features. By comparing samples produced in different labs, we show that the short-range electron-phonon coupling strength is highly sensitive to subtle processing variations. This demonstrates that a reduction of microstructural disorder through careful control of growth parameters and thermal processing can substantially improve the optoelectronic quality of selenium thin films.

\section{Results}

While the high vapor pressure and low melting point of selenium facilitate low-cost fabrication, these properties complicate many standard characterization techniques, as the material readily sublimates under moderate heating and vacuum conditions. This challenge is intensified by surface oxidation, as selenium dioxide is a far more volatile species with an even higher vapor pressure, as shown in Figure \ref{fig:Figure1}(b). To overcome these challenges, we adapted a closed-space encapsulation strategy from Nielsen et al. \cite{nielsen2024a}, depositing a thin, chemically inert film on the absorber. This overlayer effectively prevents surface oxidation and sublimation, enabling vacuum measurements and significantly expanding the accessible temperature window without compromising the structural integrity of the selenium thin film. Optical access was preserved by collecting photoluminescence and Raman signals through the transparent substrate using a microscope objective, enabling higher-sensitivity measurements than free-space optics setups.

Using this encapsulation approach, we synthesized trigonal selenium thin films following a standard process flow established for state-of-the-art devices developed by multiple research groups \cite{nakada1985a, todorov2017a, youngman2021a}; details are provided in the Methods section. The X-ray diffraction (XRD) pattern in Figure \ref{fig:Figure1}(d) confirms the polycrystalline trigonal selenium phase, with all reflections matching the ICSD reference pattern (collection code 40018) and no strong crystallographic texture observed. The top-view scanning electron microscopy (SEM), presented in Figure \ref{fig:Figure1}(e), shows the polycrystalline selenium film with grain sizes consistent with previous reports. Following the structural characterization of the selenium thin film, the device structure illustrated in Figure \ref{fig:Figure1}(c) is completed by thermally evaporating a thin gold overlayer, enabling temperature- and power-dependent Raman and photoluminescence spectroscopy.

\begin{figure}[ht!]
    \centering
    %\vspace{0.2cm}
    \includegraphics[width=\columnwidth,trim={0 0 0 0},clip]{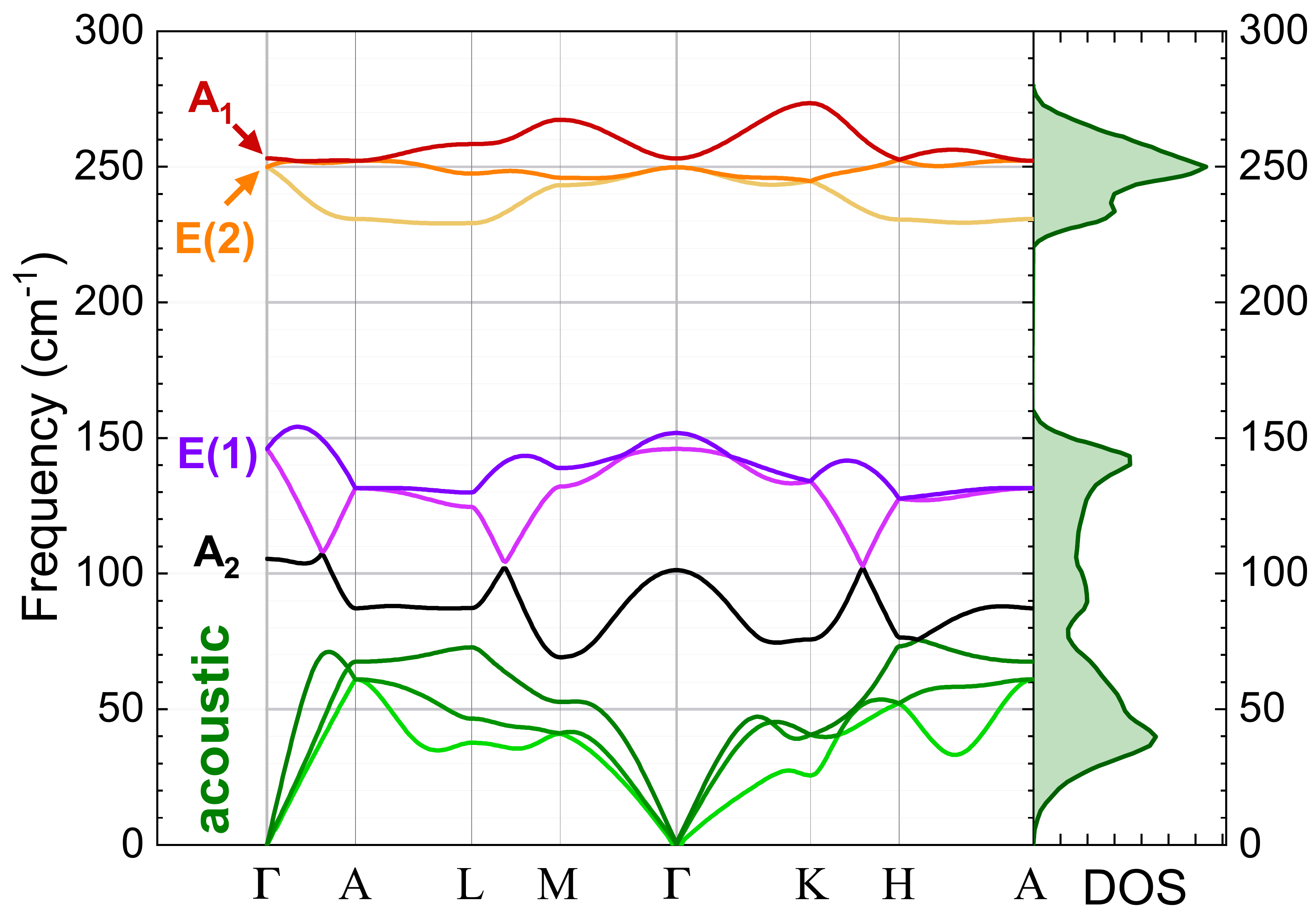}%TRIM=0 0 0 0
    \caption{DFT-calculated phonon dispersion along high-symmetry directions, alongside the total phonon density of states (DOS). Acoustic and optical modes are labeled at the $\Gamma$-point. }
    \label{fig:Figure2}
    %\vspace{-0.4cm}
\end{figure}

\subsection*{Raman spectroscopy}

To interpret the vibrational properties of the trigonal selenium thin film, we first apply group theory to identify the expected phonon modes. Trigonal selenium crystallizes in space group \textit{P}3$_\text{1}$21 (No. 152), point group 32. Group theory predicts six phonon modes with the following irreducible representation \cite{gallego2019a}:

\begin{equation}
        \Gamma_\mathrm{vib} = \mathrm{A}_1 + 2\mathrm{A}_2 + 3\mathrm{E}
\end{equation}

\noindent where (i) $\mathrm{A}_2 + \mathrm{E}$ are acoustic modes, (ii) $\mathrm{A}_2 + 2\mathrm{E}$ are infrared-active modes, and (iii) $\mathrm{A}_1 + 2\mathrm{E}$ are Raman-active modes. Here, A and E denote non-degenerate and doubly degenerate modes, respectively, with respect to rotations about the principal crystallographic axis. The E modes in trigonal selenium are sometimes labeled with numerical subscripts (e.g., E$_1$, E$_2$), but since the crystal structure is non-centrosymmetric and lacks inversion symmetry, this labeling is not appropriate. To avoid confusion with inversion parity classification, we instead label the two Raman-active E modes as E(1) and E(2).

Building on this group-theoretical framework, we next examine the phonon dispersion using first-principles density functional theory (DFT) calculations. Figure \ref{fig:Figure2} shows the phonon dispersion and the total phonon density of states (DOS), with vibrational modes at the $\Gamma$-point labeled by the symmetries introduced above. The A$_\text{1}$ and doubly degenerate E(2) modes appear nearly degenerate in frequency at $\Gamma$, suggesting they may be difficult to distinguish experimentally. The spectrum also reveals a pronounced phonon bandgap separating low- and high-frequency optical branches. While such a gap restricts available phonon scattering pathways -- potentially suppressing phonon-phonon interactions and enhancing radiative efficiency -- it may also channel energy from photoexcited carriers into specific high-energy modes, thereby increasing electron-phonon coupling and promoting non-radiative recombination. To modify or eliminate the phonon gap and thereby tailor intrinsic vibrational properties, phonon dispersion engineering through alloying, nanostructuring, or deliberate introduction of strain offers a promising strategy for enhancing optoelectronic performance potential.

\begin{figure*}[t!]
    \centering
    \includegraphics[width=\textwidth,trim={0 0 0 0},clip]{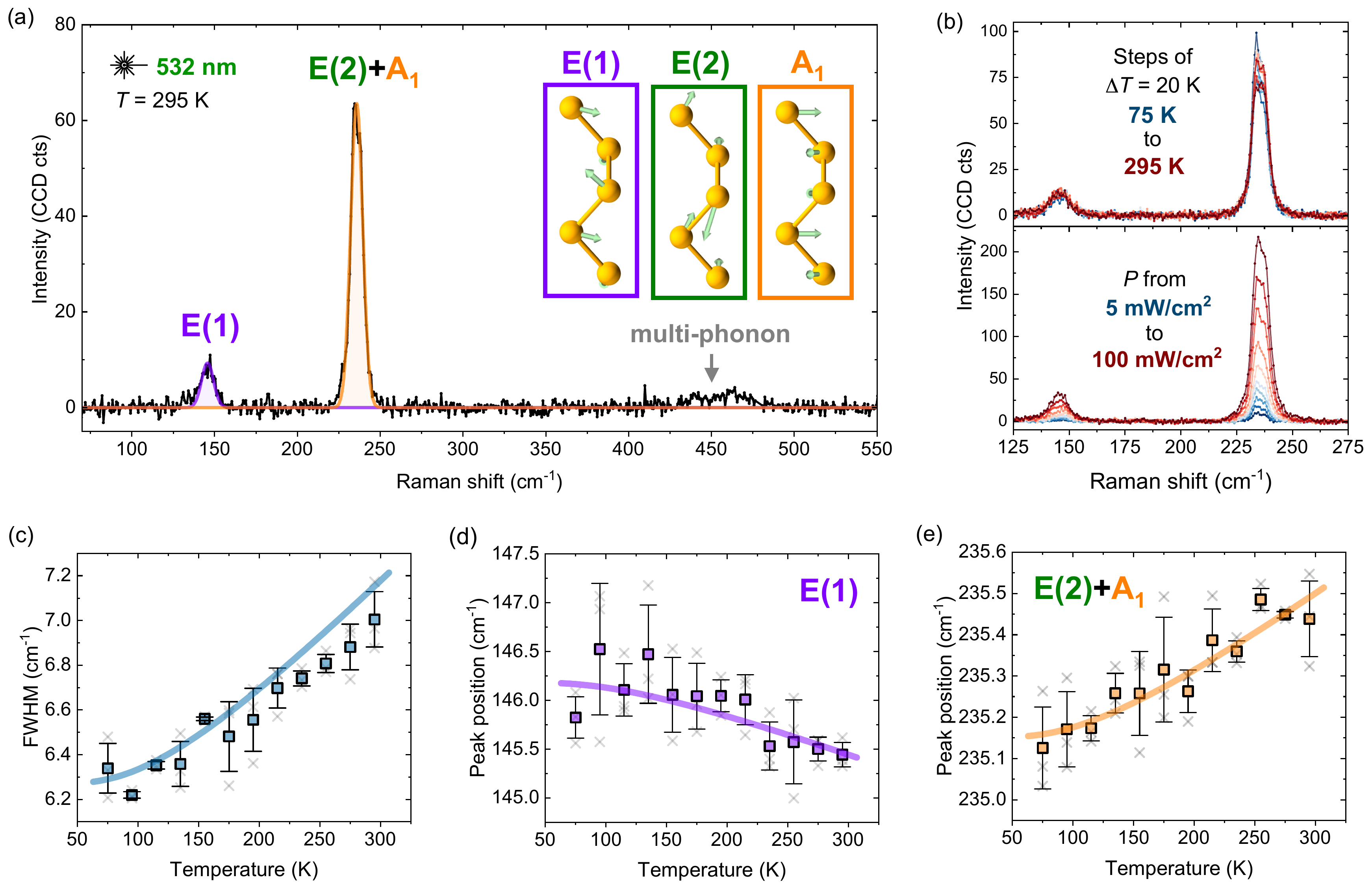}%TRIM=0 0 0 0
    \caption{(a) Deconvoluted Raman spectrum showing the E(1) and E(2)+A$_\text{1}$ modes, with the latter modeled as a single peak due to their close proximity in frequency. The insets visualize the phonon displacement patterns for all three modes. (b) Raman spectra as a function of temperature and excitation power. (c) Full width at half maximum (FWHM) of Raman peaks versus temperature, constrained to be equal for both peaks. (d) Temperature dependence of the E(1) peak position. (e) Temperature dependence of the E(2)+A$_\text{1}$ peak position. The solid lines correspond to fitted curves. The results shown in panels (c-e) are averaged from three independent temperature-dependent measurements; error bars represent one standard deviation.}
    \label{fig:Figure3}
\end{figure*}

Next, we deconvolute the experimental Raman spectrum acquired at room temperature using a rigorous methodology as detailed in Refs \cite{dimitrievska2023a, nielsen2025b}, shown in Figure \ref{fig:Figure3}(a). The low-frequency peak at $\sim$146~cm$^\text{-1}$ corresponds well with the doubly degenerate E(1) mode predicted by the phonon dispersion. The peak at $\sim$235~cm$^\text{-1}$ overlaps with both the predicted E(2) and A$_\text{1}$ modes, which are nearly degenerate and indistinguishable in energy. The slight asymmetry of this peak, particularly evident in the power-dependent Raman spectra in Figure \ref{fig:Figure3}(b), suggests contributions from two distinct but overlapping vibrational modes. However, due to the finite linewidth, limited spectral resolution, and low signal intensity, the two modes are best modeled using a single Lorentzian. These assignments are consistent with prior observations by Geick et al. \cite{geick1967a}. Additionally, a broad multi-phonon band is observed between 430 and 475~cm$^\text{-1}$, consistent with multi-phonon scattering involving 2A$_\text{1}$, 2E(2) and A$_\text{1}$ + E(2) modes. The DFT-calculated phonon displacements of the three Raman-active modes are shown as an inset in Figure \ref{fig:Figure3}(a), demonstrating that the E(1) mode is a bending type vibration that shears the covalent bonds, whereas both E(2) and A$_\text{1}$ correspond to asymmetric and symmetric bond-stretching vibrations, respectively. To better visualize these displacements, additional representations along different lattice directions are provided in Figure S1 of the Supplementary Information.

Figure \ref{fig:Figure3}(b) presents the temperature- and power-dependent Raman spectra. In addition to verifying the asymmetry of the $\sim$235~cm$^\text{-1}$ peak, which supports the interpretation of overlapping E(2) and A$_\text{1}$ modes, the power-dependent measurements are used to determine the maximum excitation power density and acquisition parameters that do not induce irreversible structural changes to the selenium thin film, thereby ensuring the integrity of subsequent Raman and photoluminescence measurements. The temperature-dependent Raman spectra are deconvoluted using the same procedure as the room-temperature spectrum in Figure \ref{fig:Figure3}(a), with the E(2) and A$_\text{1}$ modes modeled as a single Lorentzian. The full width at half maximum (FWHM) is constrained to be equal for both fitted peaks, implying similar phonon lifetimes for all single-phonon modes. The extracted FWHM and the two peak positions as a function of temperature are shown in Figures \ref{fig:Figure3}(c), (d), and (e), respectively.

To quantify the temperature dependence of the Raman peak positions and width, we fit the data using simple Bose-Einstein models, which attribute both the peak shifts and broadening to anharmonic phonon-phonon interactions. The peak positions are modeled as:

\begin{equation}\label{eq:PeakPositions}
    E(T) = E_0 - \frac{\lambda}{\exp\left(\hbar\bar{\omega} / k_\mathrm{B}T\right) - 1}
\end{equation}

\noindent where $E_0$ is the mode energy at 0\,K, $\lambda$ is the anharmonic coupling constant, $\hbar\bar{\omega}$ is the effective phonon energy, and $k_\mathrm{B}T$ is the thermal energy. The linewidth broadening is similarly modeled as:

\begin{equation}\label{eq:PeakBroadening}
    \Gamma(T) = \Gamma_0 + \frac{\Gamma_\mathrm{ph}}{\exp\left(\hbar\omega_\mathrm{ph} / k_\mathrm{B}T\right) - 1}
\end{equation}

\noindent where $\Gamma_0$ is the temperature-independent broadening, $\Gamma_\mathrm{ph}$ is the phonon-phonon coupling strength, and $\hbar\omega_\mathrm{ph}$ is the effective phonon energy. The solid lines in Figures~\ref{fig:Figure3}(c), (d), and (e) show the model fits, which reproduce the experimental trends well. The extracted fit parameters are summarized in Table~\ref{table:FittedParametersRaman}. 

\begin{table}[t!]
\small
\caption{Fitted parameters from modeling the temperature-dependent Raman modes using Equation \ref{eq:PeakPositions} and \ref{eq:PeakBroadening}, where $E_0$ is the mode energy at 0\,K, $\lambda$ is the anharmonic coupling constant, $\Gamma_0$ is the temperature-independent broadening, and $\Gamma_\mathrm{ph}$ is the phonon-phonon coupling strength. Reported uncertainties are the standard deviation errors from the fitting.}
\vspace{0.25cm}
\label{table:FittedParametersRaman}
\begin{tabular*}{0.85\columnwidth}{@{\extracolsep{\fill}}lcc} % Adjusted column alignment
    \hline \vspace{-0.25cm} \\
    Parameter \vspace{0.05cm} & E(1) & E(2)+A$_\text{1}$ \\
    \hline \vspace{-0.25cm} \\
    $E_0$ (cm$^\text{-1}$) & 146.18 $\pm$ 0.14 & 235.151 $\pm$ 0.025 \\
    $\lambda$ (cm$^\text{-1}$) & 1.6 $\pm$ 0.4 & -0.75 $\pm$ 0.07 \\
    $\Gamma_0$ (cm$^\text{-1}$) & \multicolumn{2}{c}{6.271 $\pm$ 0.011}\\
    $\Gamma_\mathrm{ph}$ (cm$^\text{-1}$) \vspace{0.05cm} & \multicolumn{2}{c}{1.96 $\pm$ 0.09} \\
    \hline
\end{tabular*}
\end{table}

As the temperature decreases, the E(1) mode is observed to blueshift. This is a vibrational fingerprint of interchain contraction, as the restoring force associated with the bond-shearing mode is strengthening, indicating an increased interchain coupling and a reduction in the distance between the selenium chains. In contrast, the E(2)+A$_\text{1}$ peak, arising from two bond-stretching modes, redshifts upon cooling -- opposite to what would be expected from simple lattice contraction. Although this redshift is smaller than the instrumental resolution, it is systematic and reproducible. This unexpected softening may result from the relaxation of grown-in compressive strain or the development of tensile strain along the chains at lower temperatures, both of which can reduce the effective restoring forces despite thermal contraction of the Se-Se bonds. Notably, this interpretation aligns with the temperature-dependent XRD measurements by Grosse et al. \cite{grosse1975a}.

\begin{figure*}[t!]
    \centering
    \includegraphics[width=\textwidth,trim={0 0 0 0},clip]{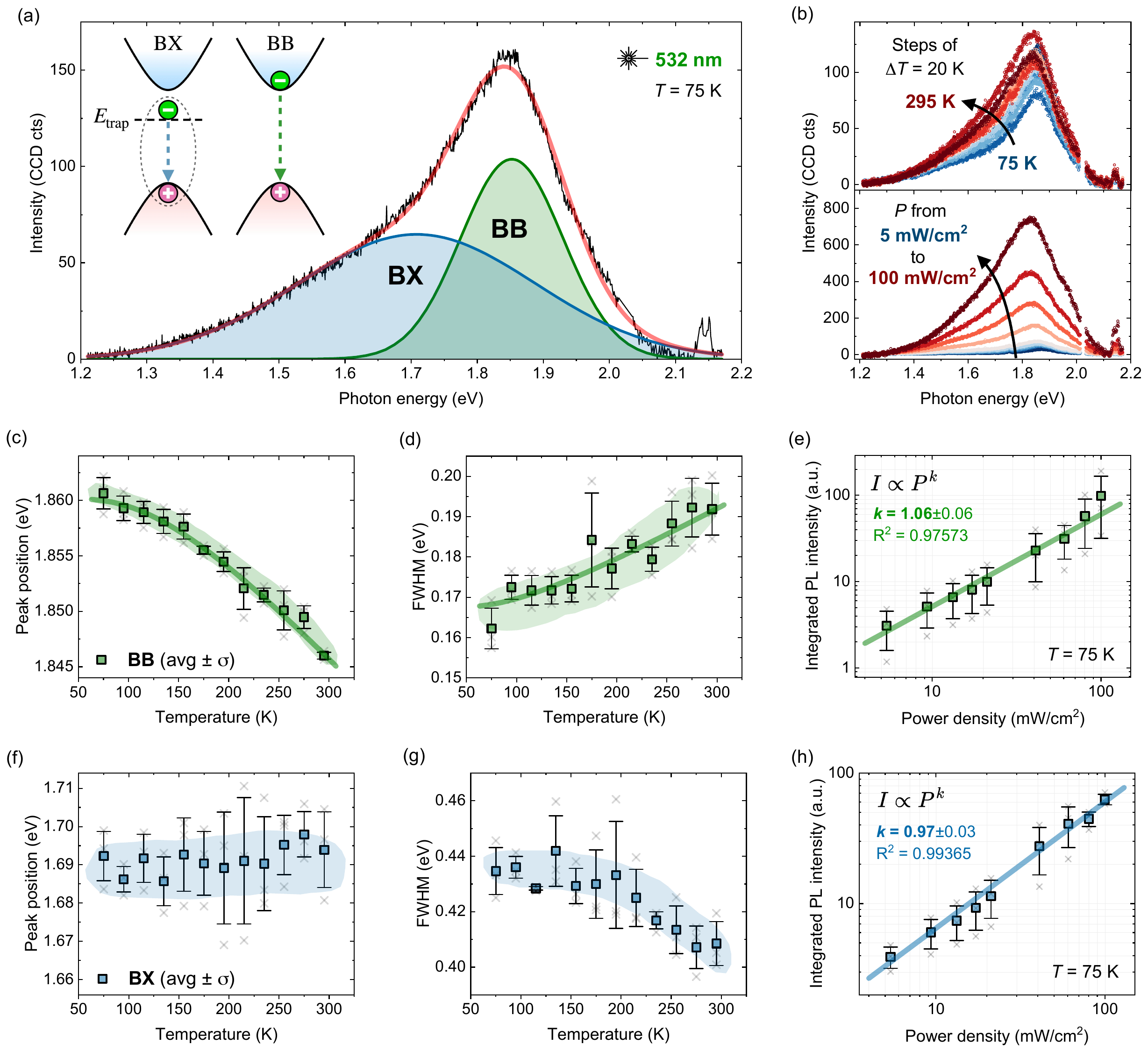}%TRIM=0 0 0 0
    \caption{Photoluminescence characterization of the trigonal selenium thin film. (a) PL spectrum measured at 75~K, deconvoluted into band-to-band (BB) and defect-bound exciton (BX) transitions, each modeled with a Gaussian peak. Inset: schematic energy band diagram illustrating the radiative recombination pathways. (b) PL spectra as a function of temperature and excitation power density. (c-e) Temperature dependence of the BB transition showing (c) peak position, (d) linewidth, and (e) integrated PL intensity as a function of excitation power at 75~K. (f-h) Corresponding data for the BX transition: (f) peak position, (g) linewidth, and (h) integrated PL intensity versus excitation power at 75~K. Solid lines in panels (c-d) and (e, h) represent fits to the Bose-Einstein models and power-law relations, respectively. Temperature-dependent fits are not applied to the BX transition since Bose-Einstein models do not adequately describe defect-bound excitons. The results shown in panels (c-h) are averaged from three independent measurements; error bars represent one standard deviation.}
    \label{fig:Figure4}
\end{figure*}

\subsection*{Photoluminescence spectroscopy}

We now turn to the optoelectronic characterization of the trigonal selenium thin film using photoluminescence (PL) spectroscopy. The spectrum measured at 75 K, shown in Figure \ref{fig:Figure4}(a), is well modeled using 2 Gaussian peaks. To determine the origin of these two emission features, we apply the same deconvolution procedure to the temperature- and excitation power-dependent spectra shown in Figure \ref{fig:Figure4}(b). For the higher-energy peak, the extracted peak position and linewidth as a function of temperature are presented in Figures \ref{fig:Figure4}(c) and (d), respectively, while the integrated PL intensity as a function of excitation power is shown in Figure \ref{fig:Figure4}(e). The corresponding results for the lower-energy peak are shown in Figures \ref{fig:Figure4}(f), (g), and (h), respectively.

The temperature dependence of the peak position and linewidth of the higher-energy emission feature, shown in Figures \ref{fig:Figure4} (c) and (d), is well described using the simple Bose-Einstein models given in Equations \ref{eq:PeakPositions} and \ref{eq:PeakBroadening}, consistent with either band-to-band or excitonic transitions. In this context, the extracted coupling strength should be interpreted as electron–phonon interactions rather than phonon-phonon interactions. In contrast, the lower-energy peak shows minimal shift in peak position and a slight narrowing of its already broad linewidth with increasing temperature. This behavior is more indicative of defect-related transitions and is not well described by simple electron-phonon coupling models. To further elucidate the nature of the two emission features, we next analyze their dependence on excitation power density at 75 K.

The excitation power dependence of the PL intensity is analyzed using the power-law relation $I \propto P^k$, where $I$ is the integrated PL intensity, $P$ is the excitation power density, and $k$ is the power-law exponent. For both emission components, we extract $k$-values close to 1. This behavior is commonly associated with band-to-band (BB) transitions, whereas defect-mediated recombination typically exhibit sub-linear behavior ($k < 1$), and excitonic transitions result in super-linear responses ($k > 1$) \cite{levanyuk1981a, schmidt1992a}. However, it is worth noting that both free-to-bound (FB) and defect-bound exciton (BX) transitions can also yield $k\approx 1$, as their transition rates depend on a single type of free carrier -- which scales linearly with excitation power density -- and on the donor or acceptor density, which is independent of the excitation intensity. The solid lines in Figures \ref{fig:Figure4}(c), (d), (e), and (h) represent the model fits, and the extracted fitting parameters are summarized in Table \ref{table:FittedParametersPL}.

\begin{table}[t!]
\small
\caption{Fitted parameters from modeling the temperature- and power-dependent PL spectra using Equations \ref{eq:PeakPositions} and \ref{eq:PeakBroadening}, where $E_0$ is the peak position at 0\,K, $\lambda$ is the electron-phonon coupling constant, $\hbar \bar{\omega}$ is the effective phonon energy, $\Gamma_0$ is the temperature-independent broadening, and $\Gamma_\mathrm{ph}$ is the electron-phonon coupling strength. Reported uncertainties are the standard deviation errors from the fitting.}
\vspace{0.25cm}
\label{table:FittedParametersPL}
\begin{tabular*}{0.85\columnwidth}{@{\extracolsep{\fill}}lcc} % Adjusted column alignment
    \hline \vspace{-0.25cm} \\
    Parameter \vspace{0.05cm} & BB & BX \\
    \hline \vspace{-0.25cm} \\
    %$E_\mathrm{g}$ (eV) & 1.8457 & 1.6981 \\
    $E_0$ (eV) & 1.860 $\pm$ 0.001 & 1.688 $\pm$ 0.003 \\
    %\lambda$ (meV) & 43 $\pm$ 16 & \textcolor{red}{-69 $\pm$ 417}   \\
    $\lambda$ (meV) & 43 $\pm$ 16 & N/A   \\
    %\hbar \bar{\omega}$ (meV) \vspace{0.05cm} & 36 $\pm$ 8 & \textcolor{red}{53 $\pm$ 131} \\
    $\hbar \bar{\omega}$ (meV) \vspace{0.05cm} & 36 $\pm$ 8 & N/A \\
    \hline \vspace{-0.25cm} \\
    $\Gamma_0$ (meV) & 167 $\pm$ 5 & 440 $\pm$ 230 \\
    %$\Gamma_\mathrm{ph}$ (meV) & 42 $\pm$ 74 & \textcolor{red}{-0.02 $\pm$ 462.25} \\
    $\Gamma_\mathrm{ph}$ (meV) & 40 $\pm$ 70 & N/A \\
    %$\hbar \omega_\mathrm{ph}$ (meV) \vspace{0.05cm} & 26 $\pm$ 30 & \textcolor{red}{0.02 $\pm$ 384.78} \\
    $\hbar \omega_\mathrm{ph}$ (meV) \vspace{0.05cm} & 30 $\pm$ 30 & N/A \\
    \hline \vspace{-0.25cm} \\
    $k$ \vspace{0.05cm} & 1.06 $\pm$ 0.06 & 0.97 $\pm$ 0.03 \\
    \hline
\end{tabular*}
\end{table}

The power-dependent PL measurements support the assignment of the higher-energy emission as a band-to-band (BB) transition, based on its nearly linear power-law exponent and its temperature evolution well described by Bose-Einstein models. The lower-energy emission exhibits a similar power-law exponent, which alone could be consistent with either a free-to-bound or defect-bound exciton (BX) transition. However, its minimal temperature-induced shift and the slight narrowing of its broad linewidth with increasing temperature are more indicative of a defect-related origin. Similar linewidth behavior has been observed for defect-bound excitons in monolayer MoS$_\text{2}$ \cite{li2020a}, and this interpretation aligns with early PL studies of vapor-phase grown selenium crystals submerged in liquid hydrogen \cite{queisser1967a}. We therefore assign the two PL features to BB and BX transitions, respectively, as schematically illustrated in the inset of Figure \ref{fig:Figure4}(a).

Finally, the characteristic phonon energy extracted from both the temperature-dependent peak shift and linewidth broadening of the BB transition closely matches that of the E(2)+A$_\text{1}$ Raman peak. This suggests that one or both of these bond-stretching vibrational modes are responsible for the dominant electron–phonon coupling driving the thermal evolution of the PL.

\section{Discussion}

The thermal evolution of the vibrational and optoelectronic parameters derived from Raman and PL spectroscopy has direct implications for the optoelectronic performance potential of the selenium thin film. As the temperature is lowered, the crystal lattice is expected to contract and the electron–phonon interactions to weaken. As a direct consequence, the optical bandgap is expected to decrease, which is consistent with our observations in Figure \ref{fig:Figure4}(c). We also observe a blueshift in the E(1) bond-shearing mode in Figure \ref{fig:Figure3}(d), suggesting a reduction in the distance between the helical chains and enhanced interchain coupling. The only anomalous response is observed in the E(2)+A$_\text{1}$ bond-stretching modes, which redshift slightly as the temperature decreases. Although this redshift is smaller than the instrumental resolution, its consistency across multiple measurements suggests a subtle softening of the restoring forces along the chain direction. This behavior could arise from the relaxation of grown-in compressive strain or the buildup of tensile stress along the chains at lower temperatures, and may contribute to the PL linewidth broadening by modifying exciton confinement potentials and/or facilitating the formation of extended structural defects. Since the crystal grains are randomly oriented, this strain cannot be attributed solely to lattice mismatch with the substrate. Instead, it likely arises from internal stress accumulated during thermal annealing and cooling during crystallization, which may promote the formation of dislocations, stacking faults, or additional grain boundaries. This interpretation is consistent with the deep-level transient spectroscopy (DLTS) study by Chen et al. \cite{chen2024a}, which attributed the dominant trap states in selenium and selenium-tellurium alloy thin films to extended, rather than point-like, defects. It also aligns with the combined computational and experimental study by Kavanagh et al. \cite{kavanagh2025a}, which showed that intrinsic point defects in trigonal selenium do not form killer recombination centers.

To evaluate how sensitive the structural quality of selenium thin films is to subtle, unintentional variations in processing, we compare samples fabricated via the standardized two-step synthesis route -- thermal evaporation of amorphous selenium followed by thermal annealing in air -- carried out in different laboratories. Specifically, we examine the thermal evolution of the PL spectra of our sample from UPC alongside a sample fabricated at DTU under nominally identical conditions. As shown in Figure S2, the DTU film features larger crystal grains, while both samples exhibit similar PL spectral shapes, including a defect-bound exciton and a band-to-band transition. Although the DTU group previously reported no detectable PL at room temperature, our measurements confirm its presence, emphasizing the effectiveness of closed-space encapsulation in preventing surface degradation and the advantage of using a microscope objective to collect PL over a larger solid angle.

A quantitative comparison of the BB peak parameters reveals that the peak position is $\approx5$ meV lower at low temperatures in the DTU sample, and the electron-phonon coupling strength is $\approx20-21$ meV, which is about half the value extracted from the UPC sample. These differences may be explained by the ultra-thin tellurium nucleation layer, which has been observed to interdiffuse into the selenium layer during the crystallization \cite{nielsen2023a}. Given that selenium–tellurium alloys exhibit a reduced optical bandgap with increasing Te content \cite{hadar2019a, zheng2022a}, the slightly lower bandgap in the DTU film may indicate greater Te incorporation, which could in turn influence lattice dynamics and weaken coupling to bond-stretching modes. Alternatively, seemingly minor differences in thermal processing -- such as marginally higher peak temperatures during the rapid thermal annealing step or variations in the cooling rate -- could also contribute to the reduced electron–phonon coupling, particularly given the observed increase in grain size. Regardless of the precise origin, these results demonstrate that the coupling strength of the electrons to the high-frequency vibrational modes is not intrinsic to the material, but depend strongly on the microstructure or the thin film and the processing history, emphasizing the need for precise control over synthesis conditions to optimize optoelectronic performance. Notably, the DTU samples have demonstrated the highest reported open-circuit voltage of 0.99 V in photovoltaic devices \cite{nielsen2022a} compared to 0.86 V for the UPC samples, highlighting the device implications in improving the optoelectronic quality of the absorber.

The interplay between vibrational and optoelectronic properties has a strong influence on overall device performance. Our results show that structural disorder in selenium thin films is not intrinsic but highly sensitive to subtle variations in processing. These extended defects likely act as killer non-radiative recombination centers. Because such defects are not necessarily radiative, their impact may not be fully captured by PL measurements alone. However, we do observe a defect-bound exciton emission feature. Prior work by Nielsen et al. \cite{nielsen2025a} identified an acceptor-like trap with a similar energy level, possibly corresponding to the (+1/0) charge transition level of selenium vacancies \cite{kavanagh2025a}. While this lower-energy exciton-related emission is unlikely to represent the primary non-radiative loss mechanism responsible for the large open-circuit voltage deficit, it may still contribute to the broad absorption onset commonly observed in both EQE and absorption spectra. This would help explain the persistent discrepancy between our PL-inferred optical bandgaps (1.84–1.86 eV), and the photovoltaic bandgaps ($\sim$1.95 eV) derived from the inflection point of the EQE \cite{todorov2017a, IvanCano2025, liu2025a, an2025a}. This discrepancy is likely a result of sub-bandgap absorption from band tailing induced by traps and/or structural disorder. Earlier work by Chen et al. \cite{chen1985a} also reported strong trapping of free excitons in selenium and proposed that such trapping could account for significant non-radiative recombination losses at higher temperatures. All of these observations are broadly consistent with the emerging picture of the microstructural and optoelectronic properties of selenium thin films.

\section{Conclusion}

In summary, we have used temperature-dependent Raman and PL spectroscopy to investigate grown-in stress, vibrational dynamics, and electron–phonon interactions in selenium thin films synthesized in different laboratories. Our results underscore the critical relation between the microstructural and optoelectronic quality, which ultimately governs device performance. We hypothesize that the structural disorder may facilitate the formation of extended defects, which are believed to be the dominant non-radiative recombination pathways limiting carrier lifetime and open-circuit voltage in selenium solar cells. A key finding is that this structural disorder is not intrinsic to the material, but is highly sensitive to subtle processing variations. Notably, films prepared using nominally similar methodologies in different labs exhibit marked differences in short-range structural disorder, which significantly influences the electron-phonon coupling strength and overall optoelectronic quality. This comparative analysis confirms that short-range electron–phonon interactions are tunable through precise control of synthesis parameters, reflecting variations in structural order and crystalline quality.

For future work, we suggest that processing strategies emphasizing careful thermal management during crystallization could help minimize anisotropic strain and suppress the formation of extended defects. Such stress may arise not only from growth conditions and substrate interactions, but also from differential thermal contraction during the cooling phase, particularly in thin films where temperature gradients and substrate anchoring can induce significant levels of grown-in stress. Post-deposition treatments, such as additional annealing protocols designed to relax residual stress and reduce structural disorder without introducing new defects, represent another promising route for improving material quality. Additionally, we demonstrate that temperature-dependent Raman and PL spectroscopy is a fast, non-destructive, and effective toolkit for assessing structural and optoelectronic quality even in randomly oriented polycrystalline films. These techniques enable the detection of strain signatures, shifts in bond-stretching and -shearing modes, and insight into the nature of radiative recombination channels. When integrated with other device-level techniques -- such as temperature-dependent current-voltage measurements, transient photoconductivity, and capacitance-based defect spectroscopy -- this multimodal approach provides a powerful platform to explore how the microscopic structure and defect landscapes influence macroscopic device performance.

\section{Methods}

\small

\subsection{Materials}\vspace{-0.35cm}

Fluorine-doped tin oxide (FTO)-coated glass substrates (TEC 15, $\sim$13~$\Omega$/sq) and tellurium pellets (99.999\%, metals basis) were purchased from Sigma-Aldrich. Selenium powder (200 mesh, 99.999\%, metals basis) was obtained from Thermo-Scientific. Gold (99.99\%) for back contacts was purchased from Neyco.

\subsection{Fabrication of selenium samples}\vspace{-0.35cm}

FTO-coated glass substrates were ultrasonically cleaned sequentially in acetone, isopropanol, and Milli-Q water, then dried using a nitrogen gun. The absorber layer was deposited using a Kenosistec co-evaporation system at a base pressure below 5$\times$10$^\text{-7}$ mbar. First, an ultrathin ($\sim$1 nm) tellurium layer was evaporated, immediately followed by the deposition of 300 nm of selenium at an evaporation rate of 10–14 nm/min, with no substrate heating. The as-deposited amorphous selenium absorber was then crystallized by annealing the samples in air on a hotplate at 200$^\circ$C (measured by the hotplate’s internal thermocouple) for 5 minutes, before being returned to vacuum. Gold back contacts were deposited in the same system at a rate of 1–2 Å/s through a shadow mask, resulting in an active cell area of 7 mm$^\text{2}$. The fabrication process for the samples from DTU, presented in the Supplementary Information, has been described in detail elsewhere \cite{nielsen2022a}.

\subsection{Raman and photoluminescence}\vspace{-0.35cm}

Raman and photoluminescence (PL) spectra were aquired using a WITec alpha300 R confocal Raman microscope in a backscattering configuration. The measurements were performed in an optical cryostat maintained at a base pressure of 10$^\text{-5}$~mbar,  with the sample chuck cooled using liquid nitrogen. The temperature of the sample chuck was monitored using a thermocouple. After the desired setpoint temperature had been reached, the system was left to stabilize for an additional 30 minutes to ensure thermal equilibrium between the sample and the stage. A 532~nm excitation laser was focused onto the sample using a 50x long working distance microscope objective (NA = 0.55), resulting in a spot diameter of approximately 1.2~$\mu$m. The backscattered light was collected and analyzed using a thermoelectrically cooled CCD spectrometer equipped with a holographic grating (150~g/mm for PL and 1800~g/mm for Raman). For both PL and Raman measurements, spectra were averaged over 100~$\mu$m line scans comprising 30 points, with an integration time of 10~s per point, to improve the signal-to-noise ratio and reduce the influence of local sample inhomogeneities. Raman peak positions were calibrated using the reference silicon peak at 520~cm$^\text{-1}$. All raw spectra and acquisition parameters are provided in the Supporting Information.

\subsection{Additional characterization}\vspace{-0.35cm}

Scanning electron microscopy (SEM) images were acquired using a Zeiss Auriga field-emission microscope operated at an acceleration voltage of 5 kV, with working distances between 3 to 5 mm. X-ray diffraction (XRD) patterns were collected using a Bruker D8 Advance diffractometer equipped with a copper X-ray tube (40~kV, 40~mA) and a Sol-X detector. Measurements were conducted in Bragg-Brentano geometry, and the detector was configured with a discriminator to suppress the Cu K$_\beta$ line and minimize fluorescence background.

\subsection{Computational details}\vspace{-0.35cm}

Calculations were performed using Density Functional Theory (DFT) through the Vienna Ab Initio Simulation Package (VASP) \cite{kresse_ab_1993,kresse_ab_1994} with the projector-augmented wave (PAW) pseudopotentials \cite{blochl_projector_1994}. The range-separated screened hybrid DFT functional of Heyd, Scuseria and Ernzerhof (HSE06) \cite{heyd_hybrid_2003} was used for atomic relaxation and displacement calculations, along with the D3 correction \cite{grimme_consistent_2010} with the zero-damping function, to account for van der Waals (vdW) type dispersion forces -- found to significantly impact the structural properties of solid-state selenium \cite{kavanagh2025a}. A plane wave energy cutoff of \SI{300}{eV} and $k$-point density of \SI{0.42}{\angstrom^{-1}} ($4\times4\times4$ for the 3-atom $P3_121$ unit cell) were used, found to give total energies converged to within \SI{1}{meV/atom} using \texttt{vaspup2.0} \cite{kavanagh_vaspup20_2023}. \texttt{phonopy} \cite{togo_first_2015} and \texttt{ThermoParser} \cite{spooner_thermoparser_2024} were used to setup and parse the phonon calculations, which used a $4\times4\times4$ supercell of the primitive trigonal selenium unit cell. The ionic dielectric response and Born effective charges were calculated using linear response theory under finite electric fields, also with the HSE06+D3 functional, in order to compute the non-analytical correction (NAC) to the phonon dispersion. See Ref \cite{kavanagh2025a} for further computational details.

\normalsize

\section*{Data availability}
All relevant data and code for analysis generated in the course of this work is freely available at doi.org/10.5281/zenodo.16639518.\\

\section*{Acknowledgements}
The work presented here is supported by the Carlsberg Foundation, grant CF24-0200. O.S.B. and E.S. acknowledge the CURIO-CITY project (PID2023-148976OB-C41); O.S.B., A.T. and M.P. acknowledge the INNO-PV project (PID2022-140226OB-C31C32); M.P. acknowledges the ENPOWER project (PCI2024-155100-2) from the CETP-Partnership Program 2023; all of them funded by MCIN/AEI/10.13039/501100011033/ FEDER. A.T. and M.P. acknowledge the SELECTRON project (CNS2023-14817) funded by MCIN/AEI/10.13039/501100011033/ NextGenerationEU/PRTR. Through our membership of the UK’s HEC Materials ChemistryConsortium, which is funded by the UK Engineering and Physical Sciences Research Council (EPSRC; EP/L000202, EP/R029431, EP/T022213), this work used the ARCHER2 UK National Supercomputing Services. We are also grateful to the UK Materials and Molecular Modelling Hub for computational resources, which is partially funded by EPSRC (EP/T022213/1, EP/W032260/1 and EP/P020194/1). S.R.K thanks the Harvard University Center for the Environment for funding a fellowship. ES acknowledge the ICREA Academia Program.

\section*{Author Contributions}
R.S.N. was responsible for data acquisition and formal analysis, wrote the original draft, contributed to funding acquisition, and co-managed the project. A.G.M. contributed equally to data acquisition. A.T. and O.S.B. synthesized the main samples and performed XRD and SEM measurements. S.R.K. developed the computational methodology. D.O.S. and A.W. conceptualized the computational aspects and contributed to funding acquisition. E.S. and M.P. supervised A.T. and O.S.B. and contributed to funding acquisition. M.D. did conceptualization, supervision, provided resources, and co-managed the project. All authors contributed to discussions and to writing – review \& editing.
%Conceptualization, Data curation, Formal analysis, Funding acquisition, Investigation, Methodology, Project administration, Resources, Software, Supervision, Validation, Visualization, Writing – original draft, Writing – review & editing

\section*{Conflicts of interest}
There are no conflicts of interest to declare.

\vfill

\newpage


\nocite{*}

\bibliography{references}% Produces the bibliography via BibTeX.

\end{document}

%\preprint{APS/123-QED}

\title{\Large{SUPPORTING INFORMATION} \\ \vspace{1cm} \large Spectroscopic Signatures of Structural Disorder and Electron-Phonon Interactions in Trigonal Selenium Thin Films for Solar Energy Harvesting}% Force line breaks with \\

\author{Rasmus S. Nielsen}
\email[]{Electronic mail: rasmus.nielsen@empa.ch}
\affiliation{Nanomaterials Spectroscopy and Imaging, Transport at Nanoscale Interfaces Laboratory, Swiss Federal Laboratories for Material Science and Technology (EMPA), Ueberlandstrasse 129, 8600 Duebendorf, Switzerland}

\author{Axel G. Medaille}
\affiliation{Universitat Politècnica de Catalunya (UPC), Photovoltaic Lab - Micro and Nano Technologies Group (MNT), Electronic Engineering Department, EEBE, Av Eduard Maristany 10-14, Barcelona 08019, Spain}
\affiliation{Universitat Politècnica de Catalunya (UPC), Barcelona Centre for Multiscale Science \& Engineering, Av Eduard Maristany 10-14, Barcelona 08019, Spain}

\author{Arnau Torrens}
\affiliation{Universitat Politècnica de Catalunya (UPC), Photovoltaic Lab - Micro and Nano Technologies Group (MNT), Electronic Engineering Department, EEBE, Av Eduard Maristany 10-14, Barcelona 08019, Spain}
\affiliation{Universitat Politècnica de Catalunya (UPC), Barcelona Centre for Multiscale Science \& Engineering, Av Eduard Maristany 10-14, Barcelona 08019, Spain}

\author{Oriol Segura-Blanch}
\affiliation{Universitat Politècnica de Catalunya (UPC), Photovoltaic Lab - Micro and Nano Technologies Group (MNT), Electronic Engineering Department, EEBE, Av Eduard Maristany 10-14, Barcelona 08019, Spain}
\affiliation{Universitat Politècnica de Catalunya (UPC), Barcelona Centre for Multiscale Science \& Engineering, Av Eduard Maristany 10-14, Barcelona 08019, Spain}

\author{Seán R. Kavanagh}
\affiliation{Harvard University Center for the Environment, Cambridge, Massachusetts 02138, United States}

\author{David O. Scanlon}
\affiliation{School of Chemistry, University of Birmingham, Birmingham B15 2TT, UK}

\author{Aron Walsh}
\affiliation{Thomas Young Centre and Department of Materials, Imperial College London, London SW7 2AZ, UK}
\affiliation{Department of Physics, Ewha Womans University, Seoul 03760, Korea}

\author{Edgardo Saucedo}
\affiliation{Universitat Politècnica de Catalunya (UPC), Photovoltaic Lab - Micro and Nano Technologies Group (MNT), Electronic Engineering Department, EEBE, Av Eduard Maristany 10-14, Barcelona 08019, Spain}
\affiliation{Universitat Politècnica de Catalunya (UPC), Barcelona Centre for Multiscale Science \& Engineering, Av Eduard Maristany 10-14, Barcelona 08019, Spain}

\author{Marcel Placidi}
\affiliation{Universitat Politècnica de Catalunya (UPC), Photovoltaic Lab - Micro and Nano Technologies Group (MNT), Electronic Engineering Department, EEBE, Av Eduard Maristany 10-14, Barcelona 08019, Spain}
\affiliation{Universitat Politècnica de Catalunya (UPC), Barcelona Centre for Multiscale Science \& Engineering, Av Eduard Maristany 10-14, Barcelona 08019, Spain}

\author{Mirjana Dimitrievska}
\email[]{Electronic mail: mirjana.dimitrievska@empa.ch}
\affiliation{Nanomaterials Spectroscopy and Imaging, Transport at Nanoscale Interfaces Laboratory, Swiss Federal Laboratories for Material Science and Technology (EMPA), Ueberlandstrasse 129, 8600 Duebendorf, Switzerland}

\maketitle

\clearpage

\begin{figure*}[ht]
    \centering
    \includegraphics[width=0.9\textwidth,trim={0 0 452 0},clip]{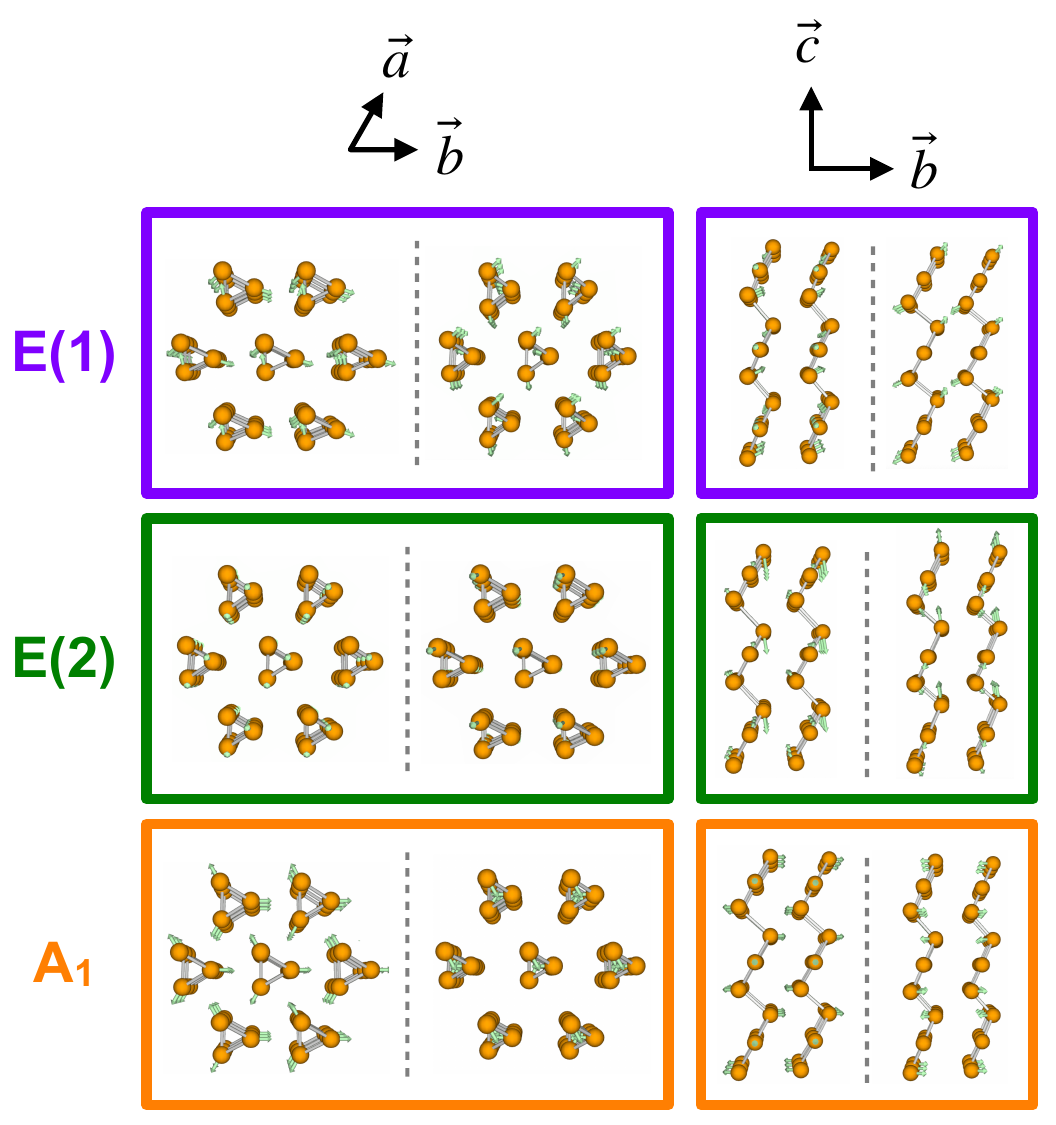}%TRIM 0 0 0 0
    \caption{Phonon displacements viewed along different lattice directions.}
    \label{fig:ESI_Figure6}
\end{figure*}

\clearpage

\begin{figure*}[ht]
    \centering
    \includegraphics[width=\textwidth,trim={0 0 0 0},clip]{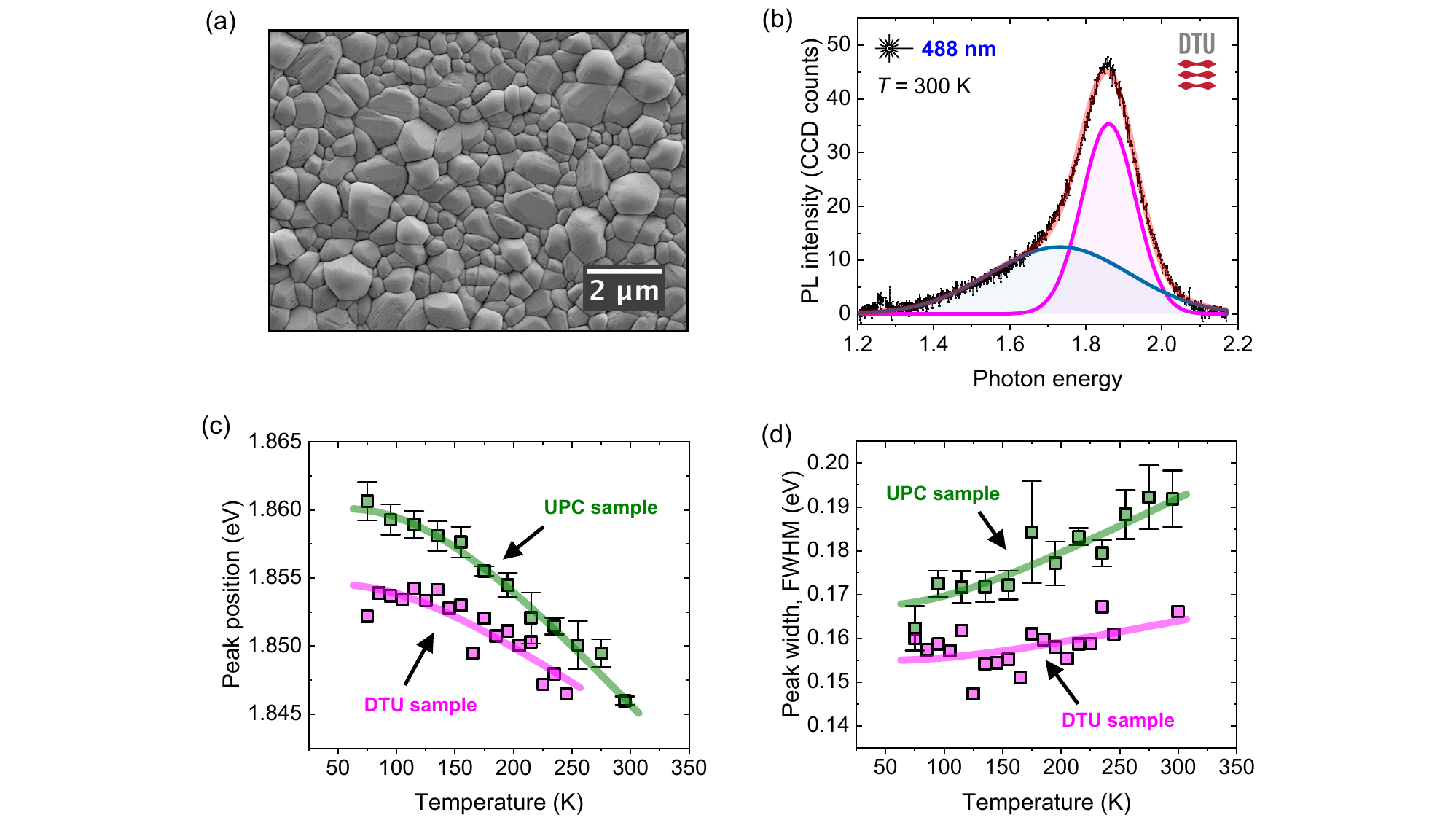}%TRIM 0 0 0 0
    \caption{Comparison of selenium thin films synthesized at DTU and UPC. (a) Top-view SEM image of the DTU sample showing larger crystal grains compared to the UPC sample. (b) Room-temperature PL spectrum of the DTU sample, revealing the same spectral features as the UPC sample -- namely, the band-to-band (BB) and defect-bound exciton (BX) transitions. However, the BB emission in the DTU sample is sharper and more dominant than the BX emission. (c) Temperature-dependence of the BB peak position for both samples. The DTU sample exhibits a slightly lower optical bandgap at low temperatures ($E_0= 1.855$~eV), but the reduced strength of electron-phonon interactions causes both samples to converge to approximately the same bandgap at room temperature. (d) Temperature-dependence of the full width at half maximum (FWHM) of the BB peak. The fitted electron-phonon coupling strengths -- $\lambda = 21.6$~meV from the peak shift and $\Gamma_\mathrm{ph} = 19.6$~meV from the peak broadening -- are approximately half the values fitted for the UPC sample. The fitted inhomogeneous broadening of the DTU sample is $\Gamma_0 = 0.155$~eV.}
    \label{fig:ESI_Figure5}
\end{figure*}

\clearpage

\begin{figure*}[ht!]
    \centering
    \includegraphics[width=\textwidth,trim={0 0 0 0},clip]{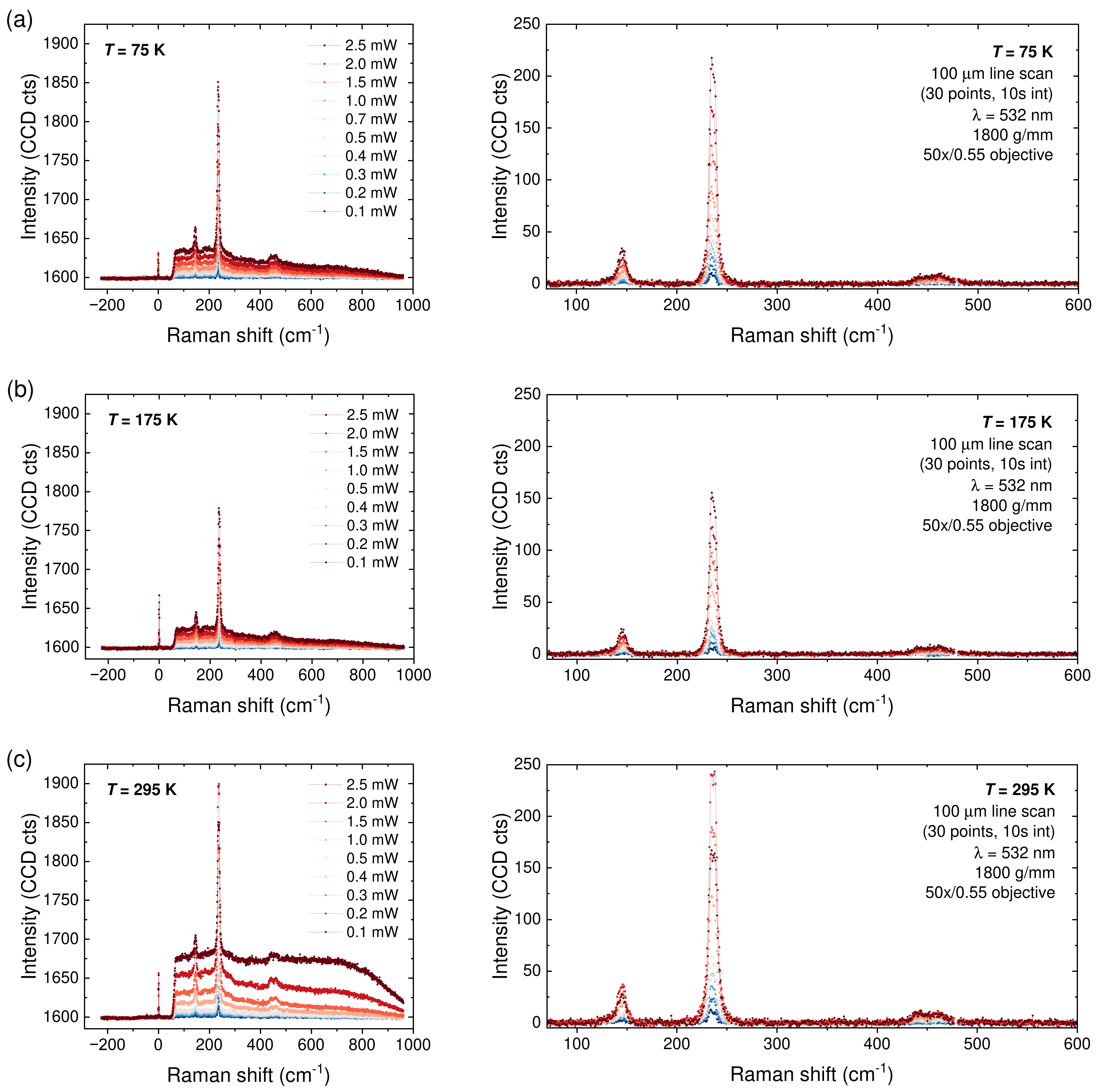}%TRIM 0 0 0 0
    \caption{Raman spectra of selenium thin films measured under varying excitation powers at (a) 75 K, (b) 175 K, and (c) 295 K. Raw spectra (left) and background-subtracted spectra (right) are shown for each temperature. Acquisition parameters are indicated in the plots.}
    \label{fig:ESI_Figure1}
\end{figure*}

\clearpage

\begin{figure*}[ht]
    \centering
    \includegraphics[width=\textwidth,trim={0 0 0 0},clip]{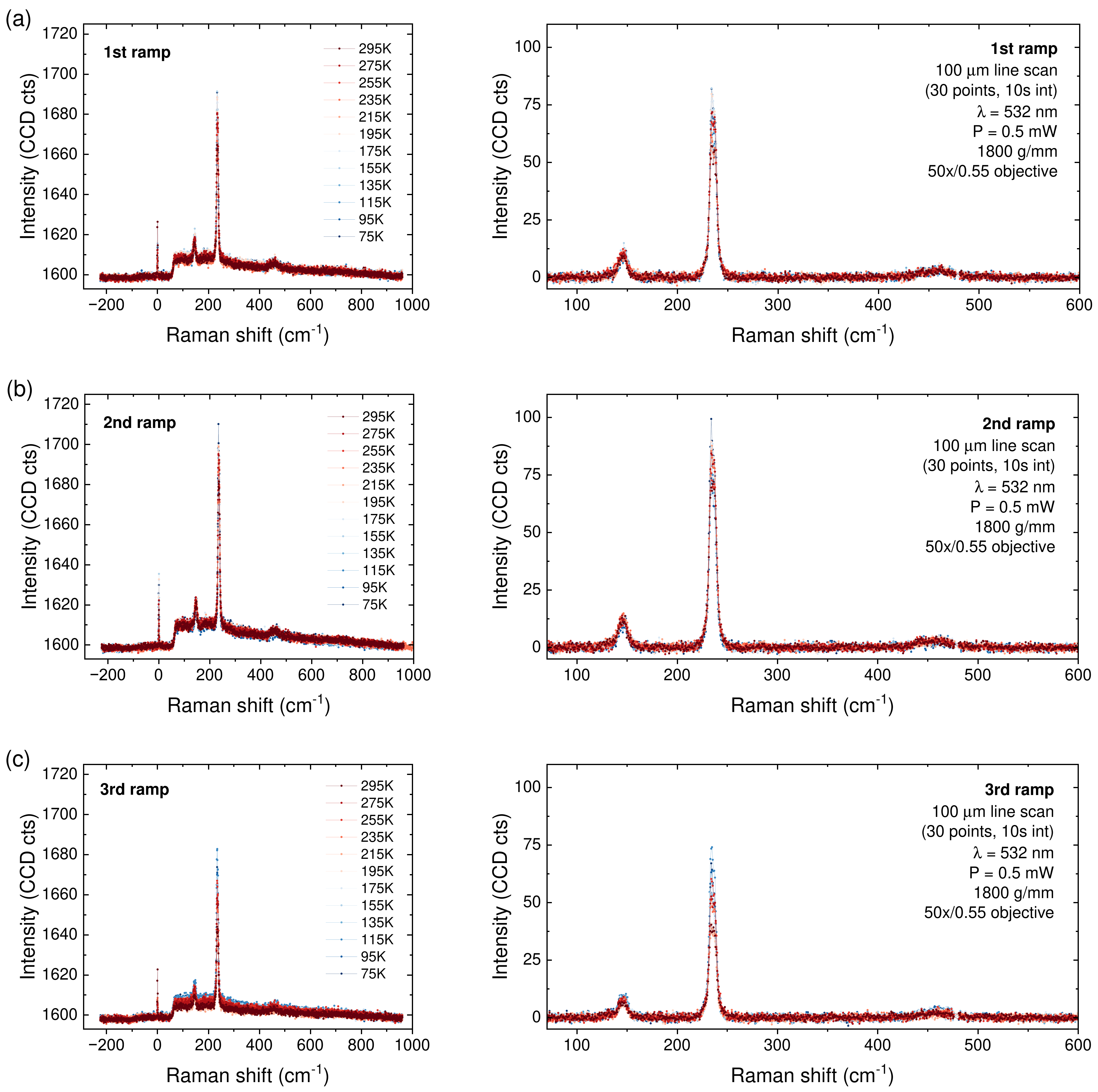}%TRIM 0 0 0 0
    \caption{Raman spectra of selenium thin films measured as a function of temperature: (a) first ramp, (b) second ramp, and (c) third ramp, each acquired at a different area on the sample. Raw spectra (left) and background-subtracted spectra (right) are shown for each measurement. Acquisition parameters are indicated in the plots.}
    \label{fig:ESI_Figure2}
\end{figure*}

\clearpage

\begin{figure*}[ht]
    \centering
    \includegraphics[width=\textwidth,trim={0 0 0 0},clip]{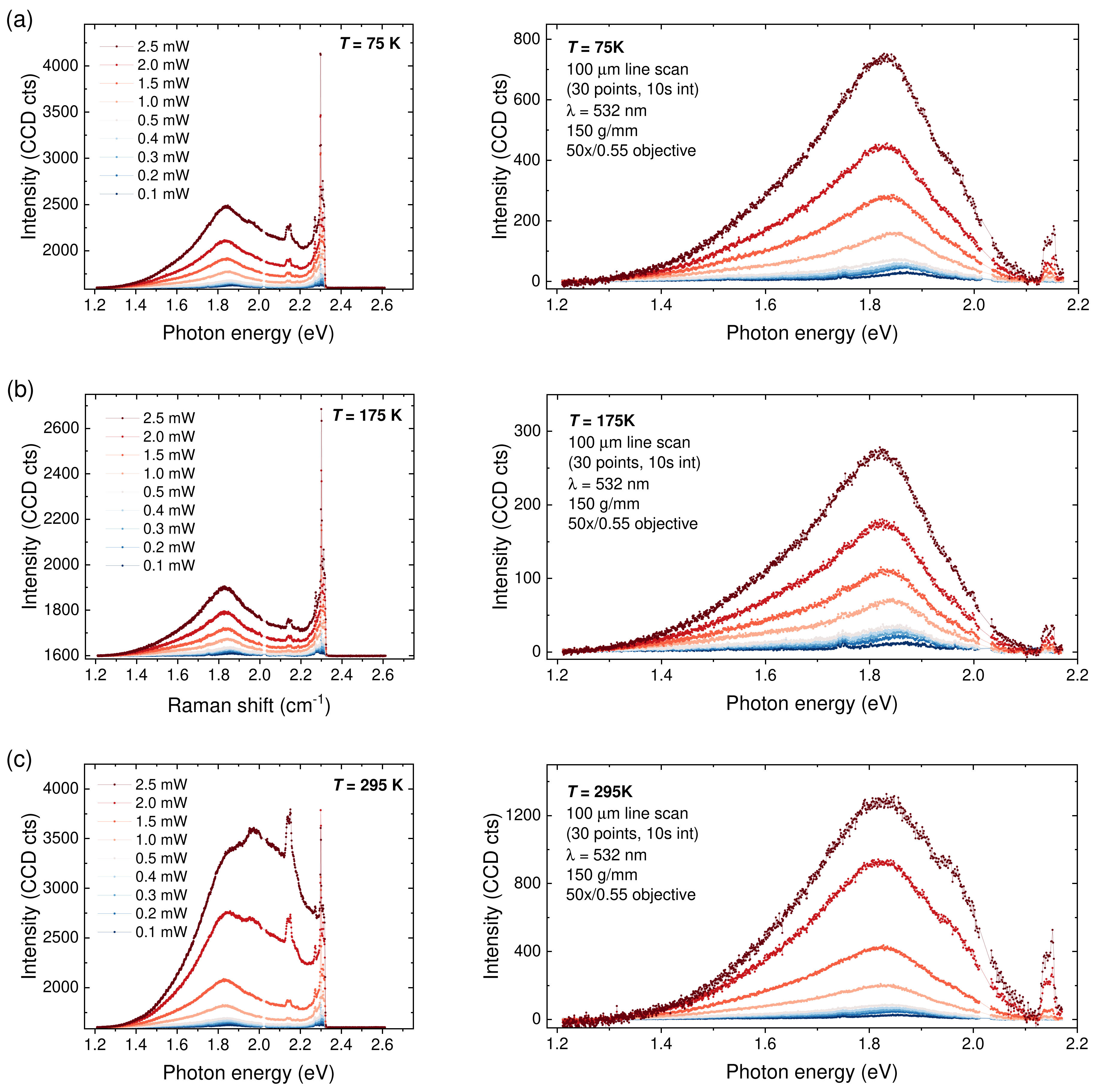}%TRIM 0 0 0 0
    \caption{Photoluminescence spectra of selenium thin films measured under varying excitation powers at (a) 75 K, (b) 175 K, and (c) 295 K. Raw spectra (left) and background-subtracted spectra (right) are shown for each temperature. Acquisition parameters are indicated in the plots.}
    \label{fig:ESI_Figure3}
\end{figure*}

\clearpage

\begin{figure*}[ht]
    \centering
    \includegraphics[width=\textwidth,trim={0 0 0 0},clip]{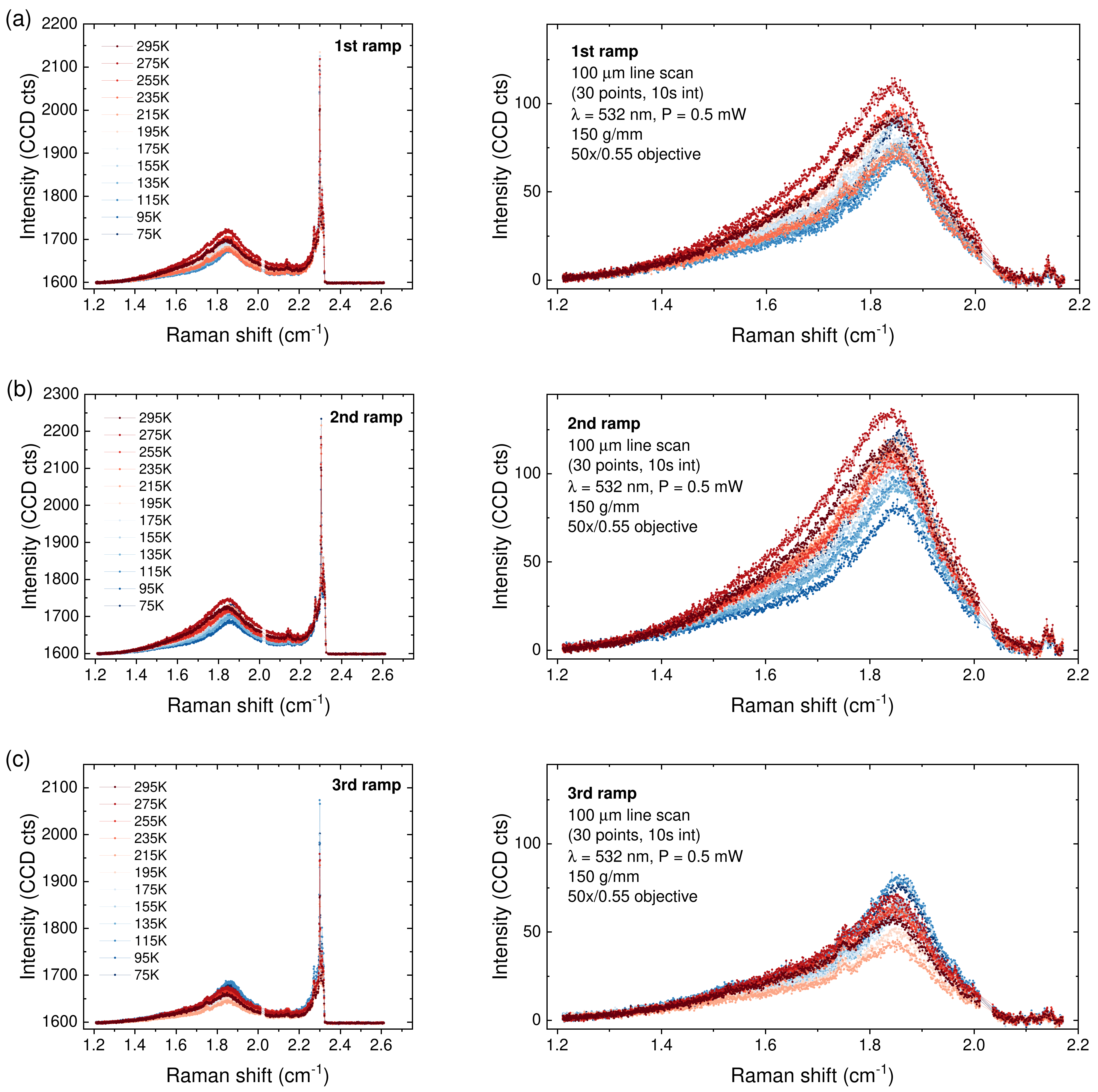}%TRIM 0 0 0 0
    \caption{Photoluminescence spectra of selenium thin films measured as a function of temperature: (a) first ramp, (b) second ramp, and (c) third ramp, each acquired at a different area on the sample. Raw spectra (left) and background-subtracted spectra (right) are shown for each measurement. Acquisition parameters are indicated in the plots.}
    \label{fig:ESI_Figure4}
\end{figure*}

\clearpage

%\nocite{*}

%\bibliography{references}% Produces the bibliography via BibTeX.